\documentclass[aps,pra,reprint,superscriptaddress,nofootinbib,longbibliography,floatfix]{revtex4-2}
\usepackage[T1]{fontenc}
\usepackage[utf8]{inputenc}
\usepackage{amsmath,amssymb,amsthm,mathtools}
\usepackage{graphicx}
\usepackage{xcolor}
\usepackage{hyperref}
\hypersetup{colorlinks=true,linkcolor=blue,citecolor=blue,urlcolor=blue}

\newtheorem{theorem}{Theorem}
\newcommand{\Hcal}{\mathcal H}
\newcommand{\Ocal}{\mathcal O}
\newcommand{\Kcal}{\mathcal K}
\newcommand{\RR}{\mathbb R}
\newcommand{\Tr}{\operatorname{Tr}}
\newcommand{\SpanR}{\operatorname{span}_{\RR}}
\newcommand{\norm}[1]{\left\lVert #1\right\rVert}
\newcommand{\abs}[1]{\left\lvert #1\right\rvert}
\newcommand{\dd}{\mathrm d}
\newcommand{\e}{\mathrm e}
\newcommand{\ii}{\mathrm i}
\newcommand{\T}{\mathsf T}

\begin{document}
\title{Observable-targeted variational quantum simulation of Hamiltonian dynamics}

\author{Leonardo Zambrano}
\email{leonardo.zambrano@icfo.eu}
\affiliation{ICFO - Institut de Ciències Fotòniques, The Barcelona Institute of Science and Technology, 08860 Castelldefels, Barcelona, Spain}

\author{Luciano Pereira}
\affiliation{ICFO - Institut de Ciències Fotòniques, The Barcelona Institute of Science and Technology, 08860 Castelldefels, Barcelona, Spain}

\author{Antonio Acín}
\affiliation{ICFO - Institut de Ciències Fotòniques, The Barcelona Institute of Science and Technology, 08860 Castelldefels, Barcelona, Spain}
\affiliation{ICREA, Passeig Lluis Companys 23, 08010 Barcelona, Spain}

\date{\today}

\begin{abstract}
Standard variational quantum simulation seeks to reproduce the evolution of the full quantum state, although many applications require only the expectation values of a few
observables. We study a variational method for pure-state Hamiltonian dynamics that updates circuit parameters to reproduce the evolution of selected expectation values. An exact error identity guides the choice of observables, motivating a construction based on repeated commutators of the target with the Hamiltonian. For Pauli observables and Pauli-rotation circuits, the update can be estimated without ancillary qubits or controlled operations for overlap estimation. Across six- and ten-qubit spin, fermionic, and molecular benchmarks, the observable-targeted update extends the median time within the target-error tolerance compared with standard variational quantum simulation with equal shot budgets. These results show that directing the variational update toward the target observable can extend accurate simulation without increasing the measurement cost per time step.
\end{abstract}
\maketitle

\section{Introduction}
\label{sec:introduction}

Quantum simulation is one of the most promising applications of quantum
processors  \cite{feynman1982simulating,lloyd1996universal}. It addresses the exponential growth of the Hilbert space with the number of constituents, which precludes an exact classical description for many-body systems beyond a few tens of particles. Current platforms already address physical questions in this regime, ranging from nonequilibrium many-body physics to quantum chemistry and materials science \cite{georgescu2014quantum, bauer2020quantum, daley2022practical}. For instance, superconducting processors and programmable atom arrays have been employed to study anomalous spin transport \cite{rosenberg2024dynamics}, thermalization and criticality \cite{andersen2025thermalization}, many-body dynamics and dynamical phase transitions \cite{bernien2017probing,zhang2017observation}, quantum phases of matter \cite{ebadi2021quantum}, and the real-time dynamics of lattice gauge theories \cite{gonzalezcuadra2025observation}.

Real-time dynamics is simulated on a digital processor by compiling the evolution operator into a gate sequence. Product formulas decompose the evolution into a sequence of local exponentials, with an error controlled by the commutators of the Hamiltonian terms \cite{lloyd1996universal,childs2021theory}, and they underlie the digital simulations carried out on current processors \cite{kim2023evidence,rosenberg2024dynamics}. Post-Trotter methods based on a linear combination of unitaries \cite{berry2015simulating}, quantum signal processing \cite{low2017optimal}, and qubitization \cite{low2019hamiltonian} achieve optimal scaling in simulated time and a logarithmic dependence of the cost on the target accuracy. These algorithms are systematically improvable and come with rigorous error bounds. Nevertheless, in all of them, the circuit depth increases with simulated time, which limits their applicability to nontrivial system sizes beyond the reach of the current generation of processors \cite{preskill2018quantum}.

An alternative approach to simulating quantum dynamics is variational quantum simulation (VQS). VQS represents the evolving state with a parameterized quantum circuit and updates its parameters using a classical processor \cite{li2017efficient,yuan2019theory}. These updates are based on the time-dependent variational principles of Dirac, Frenkel, and McLachlan \cite{dirac1930note,frenkel1934wave,mclachlan1964variational}. In McLachlan's formulation, the parameters are updated so that the instantaneous evolution of the variational state approximates the exact Hamiltonian evolution as closely as the chosen ansatz allows. Since the depth of the circuit is fixed by the ansatz and does not grow with the simulated time, VQS trades the systematic error bound of the previous algorithms for a depth that current hardware can execute, at the cost of a classical optimization and a hardware-dependent ansatz error \cite{cerezo2021variational}. Closely related formulations replace the variational equations by a sequence of fidelity-maximization problems, either step by step \cite{barison2021efficient,benedetti2021hardware} or for the whole evolution operator at once \cite{cirstoiu2020variational}. VQS has been extended to open-system and more general dynamics \cite{endo2020variational, watad2024variational} and combined with adaptive ansatz growth \cite{yao2021adaptive}. Methods for reducing measurement costs \cite{nakaji2023measurement} and bounding state errors \cite{zoufal2023error} have also been developed. The standard formulation, however, does not depend on the observable ultimately measured. The McLachlan distance weights all directions in the tangent space equally, so the available accuracy is distributed over the entire state rather than concentrated on the quantity to be predicted. Under a finite measurement budget, this is not a neutral choice, since the shots invested in estimating the variational matrices are spent on directions that may leave the target observable unaffected.

Although standard VQS uses a state-level objective, approximations tailored to selected observables have been developed in several related settings. Balian and V\'en\'eroni formulated time-dependent variational principles in which the approximation is adapted to the expectation value to be predicted \cite{balian1981time,balian1988static}. Selected observable equations have also been used in variational treatments of driven-dissipative steady states \cite{pistorius2021variational}. Observable structure also plays a central role in classical algorithms for short-time quantum dynamics \cite{wild2023classical}, sparse Pauli evolution \cite{begusic2024fast,begusic2025real}, and observable-based model reduction using Krylov operator spaces \cite{grigoletto2025exact}. Related operator-projected variational methods have been proposed for imaginary-time evolution \cite{anuar2026operator} and, recently, for real-time evolution \cite{anuar2026variational}.

In this work, we study observable-targeted variational quantum simulation, focusing on how the choice of observables affects the error in a specified target expectation value. We choose a collection of observables and update the circuit parameters so that their instantaneous evolution follows the Hamiltonian dynamics as closely as the chosen ansatz allows. This update is obtained from a least-squares problem that minimizes the mismatch in their equations of motion. When the selected observables are Pauli strings and each circuit parameter enters a single Pauli rotation, all required coefficients can be estimated through Pauli measurements on the original and parameter-shifted circuits. This avoids the ancillary qubits and controlled overlap operations used in
the Hadamard-test implementation of McLachlan evolution \cite{li2017efficient,yuan2019theory}. An exact error identity guides the choice of observables by identifying how errors in the variational dynamics affect the target expectation value, motivating a systematic construction starting from the target observable and its commutators with the Hamiltonian.

We study the performance of this proposal in six-qubit spin, Fermi-Hubbard, and molecular models, comparing it with McLachlan evolution using the same ansatz and measurement budget. The comparison measures how long the target expectation remains within a prescribed error tolerance. The observable-targeted update yields larger median reachable times across the tested models and measurement budgets, with ratios ranging from $1.06$ to $4.22$. Simulations using ten-qubit circuits extend the advantage to larger Hamiltonians, showing the same advantage with ratios from 1.24 to 11.9.

This manuscript is organized as follows: Section~\ref{sec:preliminaries} introduces the variational setting and the standard update. Section~\ref{sec:targeted-principle} develops the observable-targeted principle and its error interpretation. Section~\ref{sec:dictionaries} describes dictionary construction and measurement requirements. Section~\ref{sec:numerics} presents the numerical comparison, and Section~\ref{sec:conclusion} discusses its scope. 

\section{Standard variational quantum simulation}
\label{sec:preliminaries}

\begin{figure*}[t]
    \centering
    \includegraphics[width=\textwidth]{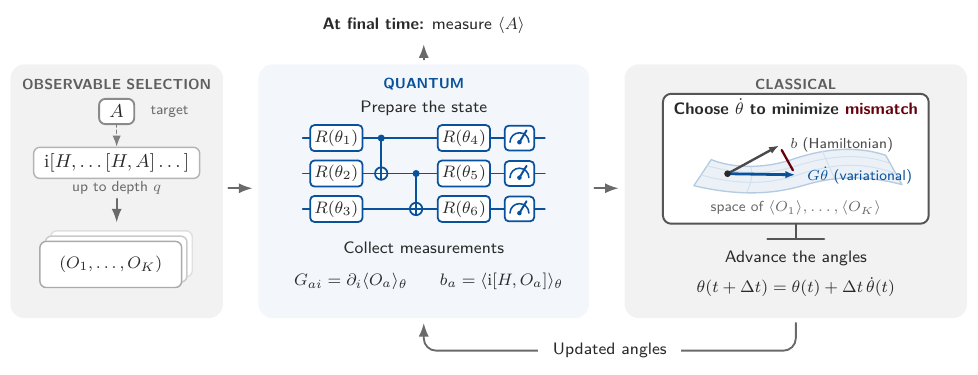}
    \caption{
    Observable-targeted variational quantum simulation.
    The Hamiltonian and target observables guide the selection
    of observables to track. The quantum processor prepares
    variational states and collects measurement data.
    The classical processor determines angle velocities that
    approximately reproduce the selected observable dynamics,
    advances the circuit angles, and returns them to the
    quantum processor. This cycle repeats until the final
    time, when the target expectation values are measured.
    }
    \label{fig:vqs-workflow}
\end{figure*}

We review standard real-time VQS \cite{li2017efficient,yuan2019theory,endo2020variational}. We first describe how a parameterized quantum circuit represents the evolving state and how McLachlan's variational principle determines the parameter updates. We then explain how these updates are implemented through quantum measurements and classical time integration. 

\subsection{Parameterized states and variational dynamics}

Let $\Hcal$ be a finite-dimensional Hilbert space and $H$ a time-independent Hamiltonian. Starting from a pure state $\lvert\psi_0\rangle\in\Hcal$, the exact state evolves according to Schr{\"o}dinger equation,
\begin{align}
    \frac{\dd}{\dd t}\lvert\psi(t)\rangle
    &=-\ii H\lvert\psi(t)\rangle.
    \label{eq:schrodinger-evolution}
\end{align}
Equivalently, the rank-one projector $\rho(t)=\lvert\psi(t)\rangle\langle\psi(t)\rvert$ obeys the von Neumann equation,
\begin{align}
    \dot\rho(t)&=\mathcal L(\rho(t)),
    \label{eq:von-neumann}
\end{align}
where $\mathcal L(X)=-\ii[H,X]$.

VQS approximates this evolution using a state prepared by a parameterized quantum circuit,
\begin{align}
    \tau_\theta
    &=\lvert\phi(\theta)\rangle\langle\phi(\theta)\rvert, 
    \label{eq:projector-ansatz}
\end{align}
where $\lvert\phi(\theta)\rangle=U(\theta)\lvert0\rangle$ and $\theta=(\theta_1,\ldots,\theta_p)$ are real parameters. We assume that the circuit depends smoothly on its parameters, which follow a trajectory $\theta(t)$ defining the approximate evolution $\tau_{\theta(t)}$. Unless stated otherwise, the initial parameters $\theta_0=\theta(0)$ are chosen to reproduce the exact initial state, $\tau_{\theta_0}=\rho(0)$.

The circuit cannot produce an arbitrary change of the state. Moving the
parameters from $\theta$ to $\theta+\epsilon v$ changes $\tau_\theta$ to first
order by
\begin{align}
    D_\theta\tau[v]
    &=\sum_{i=1}^p v_i\partial_i\tau_\theta,
    \qquad
    \partial_i=\frac{\partial}{\partial\theta_i},
    \label{eq:ansatz-differential}
\end{align}
so that $\tau_{\theta+\epsilon v}=\tau_\theta+\epsilon D_\theta\tau[v]+O(\epsilon^2)$.
Equation~\eqref{eq:ansatz-differential} therefore describes the state
derivatives that the circuit can realize at $\theta$: the $p$ operators
$\partial_i\tau_\theta$ span the accessible directions, and the
coefficients $v\in\RR^p$ specify their linear combination.
Any derivative outside this real span is unavailable to the ansatz
at $\theta$, no matter how the parameters are moved.

Simulating the dynamics requires selecting, at every instant, a combination of state derivatives the ansatz can realize. Along the variational trajectory $\sigma(t)=\tau_{\theta(t)}$, this linear combination is determined by the parameter velocity. The chain rule gives
\begin{align}
    \dot\sigma(t)
    &=D_{\theta(t)}\tau[\dot\theta(t)]
      =\sum_{i=1}^p
       \dot\theta_i(t)\,\partial_i\tau_{\theta(t)}.
    \label{eq:variational-velocity}
\end{align}
VQS therefore chooses $\dot\theta(t)$ so that this available state derivative approximates the Hamiltonian derivative $\mathcal L(\tau_{\theta(t)})$ as closely as possible according to the chosen variational objective.

\subsection{McLachlan variational dynamics}

At a fixed parameter value $\theta$, let $u\in\RR^p$ be a candidate parameter velocity. The difference between the state derivative induced by $u$ and the derivative prescribed by the Hamiltonian defines the local dynamical residual,
\begin{align}
    R_\theta(u)
    &=D_\theta\tau[u]-\mathcal L(\tau_\theta).
    \label{eq:standard-residual}
\end{align}
The residual vanishes when the circuit can reproduce the exact derivative at $\tau_\theta$. When this is not possible, the projector formulation of McLachlan's principle selects a velocity that minimizes the residual in Hilbert--Schmidt norm,
\begin{align}
    u_{\mathrm M}(\theta)
    &\in\underset{u\in\RR^p}{\arg\min}\,
      \norm{R_\theta(u)}_{\mathrm{HS}}^2,
    \label{eq:standard-projector-objective}
\end{align}
where $\norm{X}_{\mathrm{HS}}^2=\Tr(X^\dagger X)$. The selected state derivative is therefore the one available to the circuit that is closest to $\mathcal L(\tau_\theta)$ in this norm. In the numerical simulations, we regularize
the estimated update using the prescription in Appendix~\ref{app:calibration}.

Expanding the quadratic objective gives the linear system
\begin{align}
    M(\theta)u_{\mathrm M}(\theta)&=f(\theta),
    \label{eq:standard-linear-system}
\end{align}
where
\begin{align}
    M_{ij}(\theta)
    &=\Tr(\partial_i\tau_\theta\,\partial_j\tau_\theta),
    \label{eq:standard-metric}\\
    f_i(\theta)
    &=\Tr[\partial_i\tau_\theta\,\mathcal L(\tau_\theta)].
    \label{eq:standard-force}
\end{align}
The matrix $M$ is a real positive-semidefinite Gram matrix, and $f$ is commonly called the variational force. Solving Eq.~\eqref{eq:standard-linear-system} determines the parameter velocity used to evolve the circuit,
\begin{align}
    \dot\theta(t)&=u_{\mathrm M}(\theta(t)),
    \qquad
    \theta(0)=\theta_0.
    \label{eq:standard-parameter-ode}
\end{align}

The coefficients of $M$ and $f$ can also be expressed directly in terms of the state $\lvert\phi(\theta)\rangle$. Defining $\langle H\rangle_\theta=\langle\phi(\theta)\rvert H\lvert\phi(\theta)\rangle$, we obtain
\begin{align}
    \frac{M_{ij}}{2}
    &=\operatorname{Re}\!\left[
      \langle\partial_i\phi\vert\partial_j\phi\rangle
      -\langle\partial_i\phi\vert\phi\rangle
       \langle\phi\vert\partial_j\phi\rangle
      \right],
    \label{eq:phase-invariant-metric}\\
    \frac{f_i}{2}
    &=\operatorname{Im}\!\left[
      \langle\partial_i\phi\rvert H\lvert\phi\rangle
      -\langle\partial_i\phi\vert\phi\rangle
       \langle H\rangle_\theta
      \right],
    \label{eq:phase-invariant-force}
\end{align}
where all states and derivatives are evaluated at $\theta$~\cite{yuan2019theory}.

\subsection{Hybrid implementation}

Standard VQS alternates between quantum estimation and classical integration. At the current parameter value $\theta$, the quantum processor estimates the entries of $M(\theta)$ and $f(\theta)$ from Eqs.~\eqref{eq:phase-invariant-metric}~and~\eqref{eq:phase-invariant-force}. These involve overlaps and Hamiltonian matrix elements that can be estimated using Hadamard-test circuits \cite{yuan2019theory,endo2020variational}. This implementation uses an ancillary qubit and controlled circuit operations to extract the real and imaginary parts of the required overlaps.

The classical processor uses these estimates to solve Eq.~\eqref{eq:standard-linear-system} and obtain the parameter velocity. A numerical integrator then advances the parameters according to Eq.~\eqref{eq:standard-parameter-ode} \cite{li2017efficient,yuan2019theory,endo2020variational}. For example, an Euler step of size $\Delta t$ gives
\begin{align}
    \theta_{n+1}
    &=\theta_n+\Delta t\,u_{\mathrm M}(\theta_n).
    \label{eq:standard-euler-step}
\end{align}
The updated circuit prepares $\tau_{\theta_{n+1}}$, and the coefficients are estimated again whenever the integrator requires a new velocity.

The observable-targeted method retains this hybrid structure. Its defining change is the criterion used to select the parameter velocity, which we develop next.

\section{Observable-targeted variational quantum simulation}
\label{sec:targeted-principle}

Suppose that the desired output is $\Tr[A\rho(T)]$ for a
Hermitian observable $A$ and a final time $T$. Standard VQS
selects the parameter velocity by minimizing the mismatch
between the variational state derivative and the derivative
prescribed by the Hamiltonian. To tailor the update to the
desired expectation value, we first determine how this
mismatch contributes to the error in the target expectation.

\subsection{Relating the residual to the target error}

Along a differentiable variational trajectory $\sigma(t)=\tau_{\theta(t)}$, the local residual is
\begin{align}
    r(t)
    &=R_{\theta(t)}(\dot\theta(t))
      =\dot\sigma(t)-\mathcal L(\sigma(t)).
    \label{eq:residual}
\end{align}
It measures the mismatch between the variational and Hamiltonian state derivatives at each time. To relate this mismatch to the final observable error, define $\Psi_s(X)=\e^{\ii Hs}X\e^{-\ii Hs}$.

\begin{theorem}
\label{thm:target-identity}
Let $\rho(t)$ solve the von Neumann equation and let $\sigma(t)$
be a continuously differentiable variational trajectory on $[0,T]$,
with residual $r(t)$ defined in Eq.~\eqref{eq:residual}.
For every Hermitian observable $A$,
\begin{align}
    \Tr[A(\sigma(T)-\rho(T))] =&\Tr[\Psi_T(A)(\sigma(0)-\rho(0))] \nonumber\\
    &+\int_0^T
      \Tr[\Psi_{T-t}(A)r(t)]\,\dd t.
    \label{eq:target-identity}
\end{align}
\end{theorem}

The proof is given in Appendix~\ref{app:duhamel}. The first term accounts for imperfect initialization and vanishes when $\sigma(0)=\rho(0)$. The remaining term accumulates the contributions of $r(t)$ along the trajectory. At time $t$, the residual is tested against $\Psi_{T-t}(A)$, the target transported backward from the final time to $t$. The relevant family of operators is therefore the target orbit
\begin{align}
    \bigl\{\Psi_s(A):0\leq s\leq T\bigr\}.
    \label{eq:backward-orbit}
\end{align}

Residual components orthogonal to $\Psi_{T-t}(A)$ do not contribute directly to the integral at that time, even if they contribute to the state error. This observation motivates testing the residual against a finite collection of observables whose span approximates the orbit.

\subsection{Matching selected equations of motion}

We use a fixed collection of $K$ Hermitian observables,
\begin{align}
    \Ocal=(O_1,\ldots,O_K),
    \label{eq:dictionary}
\end{align}
to test the residual along the selected operator directions. We call this collection an observable dictionary and write $\SpanR\Ocal$ for its real linear span. For any Hermitian $X$, it defines the seminorm
\begin{align}
    \norm{X}_{\Ocal}^2
    &=\sum_{a=1}^K\abs{\Tr(O_aX)}^2.
    \label{eq:dictionary-seminorm}
\end{align}
This quantity measures only the components visible to the dictionary.

At a fixed parameter value $\theta$, we have
\begin{align}
    \Tr[O_aR_\theta(u)]
    &=\sum_{i=1}^p u_i\Tr(O_a\partial_i\tau_\theta)
      -\Tr[O_a\mathcal L(\tau_\theta)].
    \label{eq:coordinate-residual-derivation}
\end{align}
Writing $\langle X\rangle_\theta=\Tr(X\tau_\theta)$, we define
\begin{align}
    G_{ai}(\theta)
    &=\Tr(O_a\partial_i\tau_\theta)
      =\partial_i\langle O_a\rangle_\theta,
    \label{eq:G-definition}\\
    b_a(\theta)
    &=\Tr[O_a\mathcal L(\tau_\theta)]
      =\langle\ii[H,O_a]\rangle_\theta.
    \label{eq:b-definition}
\end{align}
Here, $G\in\RR^{K\times p}$ encodes how the circuit parameters affect the selected expectation values, while $b\in\RR^K$ contains the Hamiltonian derivatives at the current variational state. Consequently, we have
\begin{align}
    \Tr[O_aR_\theta(u)]
    &=(G(\theta)u-b(\theta))_a,
    \label{eq:coordinate-residual}
\end{align}
so the total squared mismatch in the selected operator directions is
\begin{align}
    \norm{R_\theta(u)}_{\Ocal}^2
    &=\norm{G(\theta)u-b(\theta)}_2^2.
    \label{eq:coordinate-residual-seminorm}
\end{align}

These coefficients have a direct dynamical interpretation. For a parameter velocity $u$, the induced rate of change of $\langle O_a\rangle_\theta$ is $(Gu)_a$, whereas Hamiltonian evolution from the same state gives the rate $b_a$. Thus, each entry of $Gu-b$ is exactly the mismatch in one of the selected instantaneous equations of motion.

We define the observable-targeted velocity by the unweighted
least-squares objective
\begin{align}
    u_\star(\theta)
    &\in \underset{u\in\RR^p}{\arg\min}\,
      \norm{G(\theta)u-b(\theta)}_2^2.
    \label{eq:targeted-objective}
\end{align}
The observable-targeted update minimizes the same residual as the McLachlan evolution, but uses the dictionary seminorm instead of the Hilbert--Schmidt norm. The two objectives coincide when the dictionary is a complete Hilbert--Schmidt orthonormal operator basis. A restricted dictionary penalizes only the mismatch in the selected equations of motion, without requiring the remaining residual components to be small.

For noisy or ill-conditioned data, we minimize
\begin{align}
    u_\lambda(\theta)
    &=\underset{u\in\RR^p}{\arg\min}\,
      \bigl\{\norm{Gu-b}_2^2+\lambda\norm{u}_2^2\bigr\},
    \label{eq:ridge-objective}
\end{align}
with $\lambda>0$. The regularized objective has a unique minimizer and balances equation matching against the size of the parameter update. Solver details are given in Appendix~\ref{app:implementation}.

\subsection{Quantum simulation algorithm}

With exact initialization, Eq.~\eqref{eq:target-identity} shows when matching the selected equations is enough to recover the exact final expectation value. Suppose that every operator in the target orbit can be written as a linear combination of dictionary observables, possibly plus a multiple of the identity. If the variational trajectory satisfies all selected equations of motion exactly then $\Tr[O_a r(t)]=0$ for every $a$ and every time $t$.
Since $\Tr r(t)=0$ as well, the integrand in Eq.~\eqref{eq:target-identity} vanishes, giving zero target error.

When the dictionary only approximates the target orbit, matching its equations does not control the contribution from operator directions outside its span. A small mismatch in the selected equations therefore does not, by itself,
guarantee a small target error.

For a target observable $A$ and time horizon $T$, the simulation proceeds as follows.
\begin{enumerate}
    \item \emph{Choose and initialize the ansatz.}
    Select a differentiable circuit $U(\theta)$ and initialize its parameters so that $\tau_{\theta_0}=\rho(0)$.

    \item \emph{Construct the dictionary and fix the numerical
    prescription.}
    Choose $\Ocal$ so that its span approximates the target orbit over $[0,T]$. Repeated commutators of $A$ with $H$ provide a systematic starting point, as described in Sec.~\ref{sec:dictionaries}. Fix the dictionary normalization, regularization prescription, and classical time integrator.

    \item \emph{Estimate the local coefficients.}
    At each integrator stage, estimate $G$ and $b$ at the current parameter value using Eqs.~\eqref{eq:G-definition} and \eqref{eq:b-definition}. For Pauli dictionaries and circuits admitting parameter-shift rules, this requires only Pauli measurements on the original and parameter-shifted circuits, without measurement ancillas.

    \item \emph{Compute the parameter velocity.}
    Solve the least-squares problem in Eq.~\eqref{eq:targeted-objective}, or its regularized version in Eq.~\eqref{eq:ridge-objective}, using the estimated coefficients.

    \item \emph{Propagate in time.}
    Update the circuit parameters using the computed velocity. For example, an Euler step of size $\Delta t$ gives
    \begin{align}
        \theta_{n+1}
        &=\theta_n+\Delta t\,u(\theta_n),
    \end{align}
    where $u$ is the selected velocity. Estimate the coefficients again at the updated parameters and repeat until the final time $T$. 

    \item \emph{Measure the target.}
    At the final parameter value, estimate $\Tr[A\tau_{\theta(T)}]$ as the approximation to $\Tr[A\rho(T)]$.
\end{enumerate}

The target-error identity describes a differentiable trajectory. In practice, estimating the coefficients and integrating the parameter equations numerically introduce sampling and time-discretization errors.

\section{Constructing and measuring target-adapted dictionaries}
\label{sec:dictionaries}

The error identity identifies which operator directions matter, but computing the full target orbit can itself be difficult. We therefore construct finite dictionaries from the Hamiltonian and target, using nested commutators as a systematic starting point.

\subsection{Heisenberg Krylov spaces and Pauli dictionaries}

Define the Heisenberg generator $\delta_H(X)=\ii[H,X]$. The order-$q$ real Krylov space generated by $A$ is
\begin{align}
 \Kcal_q^H(A)&=\SpanR\{A,\delta_H(A),\ldots,\delta_H^q(A)\}.
 \label{eq:krylov-space}
\end{align}
Its connection to the target orbit follows from
\begin{align}
 \Psi_s(A)&=\e^{s\delta_H}(A)
          =\sum_{m=0}^\infty\frac{s^m}{m!}\delta_H^m(A).
 \label{eq:heisenberg-series}
\end{align}

For qubit systems, write
\begin{align}
 H&=\sum_\ell h_\ell P_\ell, &
 A&=\sum_m a_m Q_m,
 \label{eq:pauli-expansions}
\end{align}
where $P_\ell$ and $Q_m$ are Pauli strings. Each nonzero $\ii[P_\ell,Q_m]$ is a real multiple of another Pauli string. Starting from the strings in $A$, we recursively collect the distinct strings reached through at most $q$ such commutators. Denote the resulting set by $\mathsf S_q(A)$. Before any additional truncation,
\begin{align}
 \Kcal_q^H(A)&\subseteq\SpanR\mathsf S_q(A).
 \label{eq:pauli-krylov-inclusion}
\end{align}
The inclusion can be strict: matching each constituent Pauli string imposes more equations than matching the nested commutators as whole operators. The numerical study uses Pauli dictionaries, with each distinct nonidentity string included once and normalized to have eigenvalues $\pm1$. Any identity component is omitted because its pairing with the residual vanishes.

For several targets, one may use a union of their dictionaries and propagate a common variational trajectory. This remains practical when the union can be constructed and measured at an acceptable cost.

\subsection{Measurement requirements}

The observable-targeted update requires the expectation-value derivatives $G_{ai}(\theta)=\partial_i\langle O_a\rangle_\theta$ and the Hamiltonian derivatives $b_a(\theta)=\langle\ii[H,O_a]\rangle_\theta$. Both can be estimated using measurements of Pauli observables when the dictionary consists of Pauli strings.

To estimate $G$, suppose that $\theta_i$ appears in a single circuit gate of the form
\begin{align}
    U_i(\theta_i)
    &=\exp(-\ii\theta_i P_i/2),
\end{align}
where $ P_i=P_i^\dagger$ and $P_i^2=I$. The parameter-shift rule then gives \cite{schuld2019evaluating}
\begin{align}
    G_{ai}(\theta)
    &=\frac{1}{2}\left[
      \langle O_a\rangle_{\theta+\frac{\pi}{2}e_i}
      -\langle O_a\rangle_{\theta-\frac{\pi}{2}e_i}
      \right],
    \label{eq:parameter-shift}
\end{align}
where $e_i$ is the $i$th coordinate vector. Each derivative is therefore obtained by measuring the same observable on two circuits with shifted parameter values. 

To estimate $b$, expand the commutator using the Pauli decomposition of the Hamiltonian,
\begin{align}
    \ii[H,O_a]
    &=\sum_\ell h_\ell\,\ii[P_\ell,O_a]
      =\sum_P c_{aP}P.
    \label{eq:force-pauli}
\end{align}
It follows that
\begin{align}
    b_a(\theta)
    &=\sum_P c_{aP}\langle P\rangle_\theta.
    \label{eq:force-estimation}
\end{align}
These expectations are measured on the unshifted circuit.

Thus, for these circuit gates and Pauli dictionaries, all coefficients can be estimated using Pauli measurements on the original and parameter-shifted circuits, without ancillary qubits or controlled operations for overlap estimation. Compatible observables can be grouped into common measurement settings, and shot-allocation methods can be applied to the resulting estimators \cite{crawford2021efficient}. Further implementation details are given in Appendix~\ref{app:implementation}.

The observable-targeted system contains $Kp$ derivative coefficients
and $K$ entries of $b$, compared with $p(p+1)/2$ independent entries of $M$ and $p$ entries of $f$ in standard projector-level VQS. For $K\ll p$, the observable-targeted
update therefore requires fewer coefficients, offering a
potential reduction in measurement cost. The achievable
savings depend on measurement grouping, estimator variances,
circuit depth, and the number of time-integrator stages.

Dictionary size, therefore, creates a practical trade-off. Adding observables can improve coverage of the target orbit, but also introduces more equations for the ansatz to match and potentially more quantities to measure. A useful dictionary must balance these requirements over the times of interest.

\section{Numerical comparison under finite measurement statistics}
\label{sec:numerics}

\begin{figure*}[t]
 \centering
   \includegraphics[width=\textwidth]{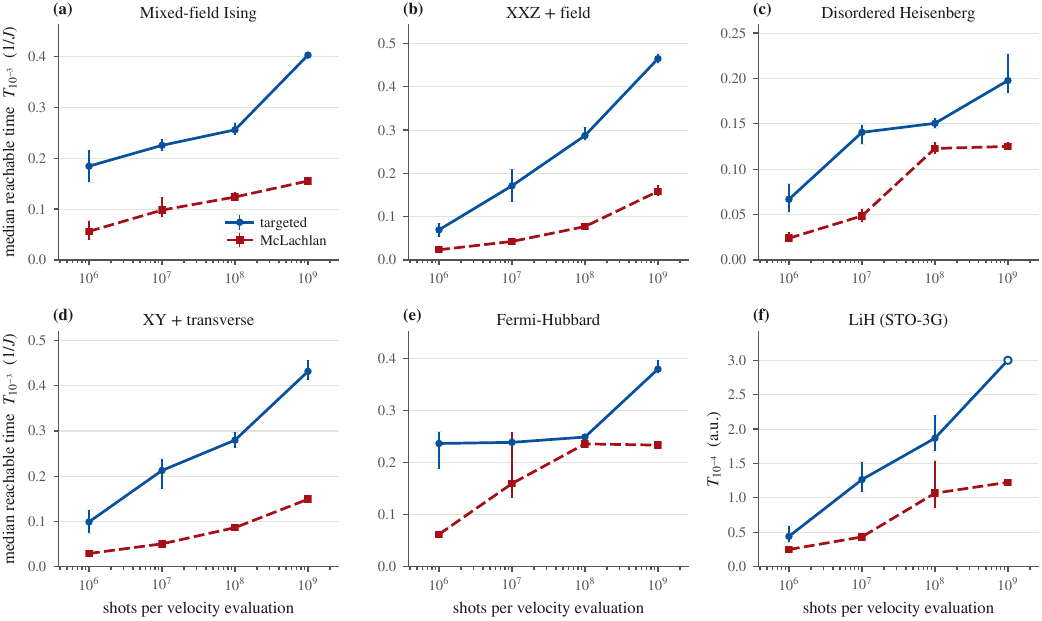}
\caption{Observable dynamics under finite measurement statistics with six qubits. Median reachable time $T_\varepsilon$ versus shots per velocity evaluation for (a) mixed-field Ising, (b) XXZ, (c) disordered Heisenberg, (d) XY, (e) Fermi--Hubbard, and (f) LiH. The tolerance is $\varepsilon=10^{-3}$ in (a)--(e) and $\varepsilon=10^{-4}$ in (f). Blue and dark red denote observable-targeted and McLachlan evolution, respectively. Both methods use six qubits and two circuit layers. Error bars show the interquartile range over 100 independent measurement realizations. The open marker for observable-targeted LiH evolution at $S=10^9$ indicates a lower bound: 85 of 100 trajectories remain within the tolerance through $T_{\max}=3$, so the median and both quartiles lie beyond the observation window. Time is measured in inverse lattice energy units in (a)--(e) and atomic units $\hbar/E_{\mathrm h}$ in (f), where $E_{\mathrm h}$ is the Hartree energy. The target expectation is evaluated exactly along each sampled trajectory.}
 \label{fig:numerics}
\end{figure*}

We investigate whether observable targeting improves the dynamics obtained from a shallow circuit in the presence of finite measurement statistics. Within each model, the observable-targeted and McLachlan methods use the same ansatz and measurement expenditure per unit of simulated time. Appendix~\ref{app:numerical-protocol} details the protocol.

\subsection{Simulated models and accuracy metric}

We perform numerical simulations of several Hamiltonians commonly used as benchmarks for quantum simulation, including a mixed-field Ising chain, an XXZ chain, a Heisenberg chain with one fixed realization of random longitudinal fields, an anisotropic XY chain, a three-site Fermi-Hubbard model, and an active-space Hamiltonian for LiH. Each system is represented by six and ten qubits. The spin-chain target is the central correlation $Z_2Z_3$, the Hubbard target is the central double occupancy $D_1=n_{1\uparrow}n_{1\downarrow}$, and the molecular target is the total occupation $N_1=n_2+n_3$ of an initially doubly occupied excited spatial orbital. Here, $Z_j$ is the Pauli $Z$ operator on qubit $j$, and $n_{j\sigma}$ is the fermion number operator at site $j$ with spin $\sigma\in\{\uparrow,\downarrow\}$. Both methods use two layers of Hamiltonian-inspired Pauli rotations. The observable-targeted dictionary is generated with commutator depth $q=3$ for six qubits and $q=2$ for ten qubits.

At each velocity evaluation, a total of $S$ measurement shots is distributed among the measurement settings required by the corresponding method. Both methods use Heun integration \cite{hairer1993} with step $\Delta t=5\times10^{-3}$ for six qubits and $2.5\times10^{-3}$ for ten qubits. Each time step requires two velocity evaluations and therefore uses $2S$ shots. With the same $S$ and $\Delta t$, both methods have the same measurement expenditure per unit simulated time. This comparison fixes the number of shots, rather than the circuit execution time.

We measure accuracy by the first time the target error exceeds a prescribed tolerance,
\begin{align}
 T_\varepsilon=\inf\{t\geq0:
 |\Tr[A(\sigma(t)-\rho(t))]|>\varepsilon\}.
 \label{eq:reachable-time}
\end{align}
Thus, the benchmark asks how long each method maintains accurate predictions along the trajectory, rather than only at a single final time. We set $\varepsilon=10^{-3}$ for the lattice models and $\varepsilon=10^{-4}$ for LiH in six qubits, and $\varepsilon=10^{-3}$ for all models in ten qubits. The tighter molecular tolerance accounts for the smaller variation of its target: the orbital population changes by approximately $0.0715$ over the simulated interval, compared with a median variation of $0.82$ for the lattice targets. The chosen tolerances therefore impose comparable accuracy relative to these variations.

The target expectation is evaluated exactly along each sampled trajectory. For each model, method, and budget $S\in\{10^6,10^7,10^8,10^9\}$, we use 100 independent measurement realizations and report the median reachable time and interquartile range. Trajectories that remain within the tolerance through $T_{\max}=3$ provide lower bounds on their crossing times. More details are in Appendix~\ref{app:numerical-protocol}.

\subsection{Observable accuracy across models}

Figure~\ref{fig:numerics} compares the reachable times as the measurement budget increases. The observable-targeted method gives a larger median in every comparison. For the Ising, XXZ, and XY chains, the ratio of observable-targeted to McLachlan medians ranges from $2.07$ to $4.22$ over the four budgets. The largest ratio occurs for the XY chain at $S=10^7$, where the median reachable time increases from $0.0504$ to $0.213$. For the disordered Heisenberg chain, the ratios range from $1.23$ to $2.90$.

The Fermi-Hubbard results show a stronger dependence on the measurement budget. At $S=10^6$, the median increases from $0.0620$ to $0.237$, a factor of $3.81$. At $S=10^8$, the improvement ratio is $1.06$, from $0.236$ to $0.249$. At $S=10^7$, the ratio of medians is $1.50$, but its estimated uncertainty interval, $[0.97,1.62]$, includes unity. 

The molecular example involves a substantially larger measurement problem within the same six-qubit space. Its Hamiltonian contains 61 non-identity Pauli terms, yielding 122 circuit parameters and 21\,805 observable-targeted measurement settings per velocity evaluation, compared with 14\,966 for the McLachlan evolution. At $S=10^7$, the median reachable time increases from $0.431$ to $1.27$ atomic units. At $S=10^9$, 85 of 100 observable-targeted trajectories remain within the tolerance through $T_{\max}=3$, so the observable-targeted median is bounded below by $3$. Compared with the McLachlan median of $1.23$, this gives a ratio of at least $2.45$.

\begin{figure*}[t]
 \centering
   \includegraphics[width=\textwidth]{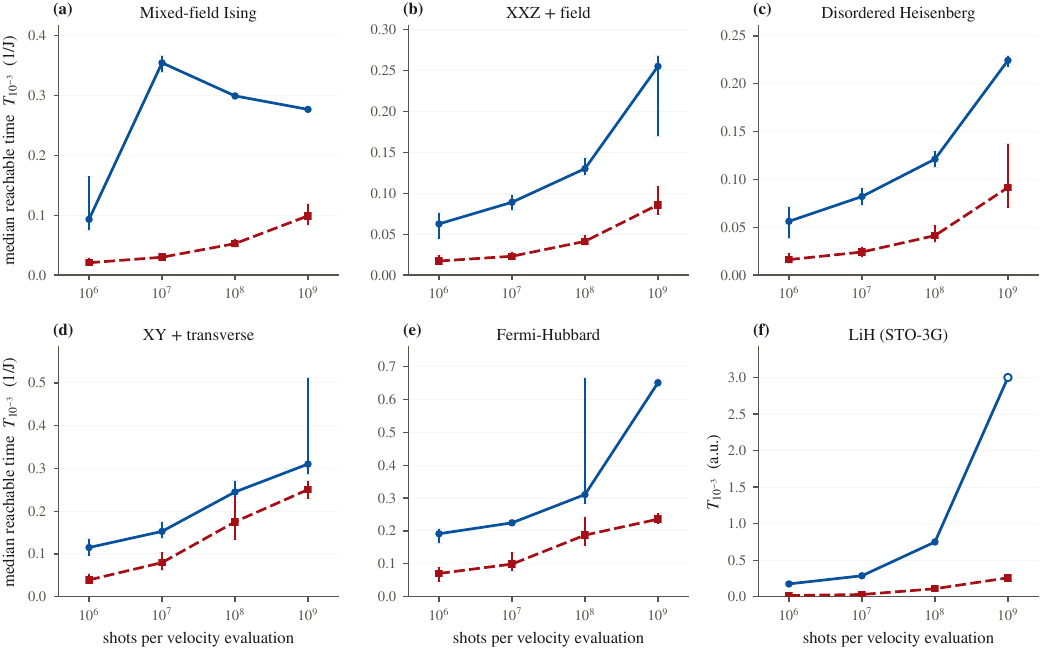}
\caption{Observable dynamics under finite measurement statistics with ten qubits. Median reachable time $T_\varepsilon$ versus shots per velocity evaluation for (a) mixed-field Ising, (b) XXZ, (c) disordered Heisenberg, (d) XY, (e) Fermi--Hubbard, and (f) LiH. The tolerance is $\varepsilon=10^{-3}$ in all panels. Blue and dark red denote observable-targeted and McLachlan evolution, respectively. Both methods use two circuit layers. Error bars show the interquartile range over 100 independent measurement realizations. Time is measured in inverse lattice energy units in (a)--(e) and atomic units $\hbar/E_{\mathrm h}$ in (f), where $E_{\mathrm h}$ is the Hartree energy. The target expectation is evaluated exactly along each sampled trajectory.}
 \label{fig:ten-qubits}
\end{figure*}

Figure~\ref{fig:ten-qubits} shows the benchmarks with ten qubits, where the circuits contain roughly twice as many parameters. We consider a dictionary depth of $q=2$. Both methods use the budget-dependent regularization for the six-qubit simulations, with a smaller observable-targeted prefactor (see Appendix~\ref{app:calibration}). The observable-targeted method yields a larger median reachable time across all comparisons. For the lattice models, the ratios range from $1.24$ for the XY chain at $S=10^9$ to $11.9$ for the Ising chain at $S=10^7$. For LiH, they range from $6.8$ to $9.2$ for $S\leq10^8$, and at $S=10^9$ the observable-targeted median is bounded below by $T_{\max}$, giving a ratio of at least $11.9$. The advantage therefore persists for larger systems, even with a shallower dictionary.

For the Ising chain, the observable-targeted median peaks at $S=10^7$ and decreases at larger budgets. In this case, the regularized update computed from noisy coefficients is biased, and at intermediate budgets this bias partially cancels the systematic error of the ansatz and dictionary. The peak therefore reflects a cancellation of errors rather than a more accurate description of the dynamics.

\section{Conclusion}
\label{sec:conclusion}

Observable-targeted VQS updates a quantum circuit to reproduce the dynamics of selected observables rather than the full state. With exact initialization, Theorem~\ref{thm:target-identity} expresses the target error as the accumulated contribution of the residual, paired with the target transported backward in time. This identity guides the choice of observables: repeated commutators of the target with the Hamiltonian provide a systematic way to construct the dictionary. For Pauli dictionaries and circuits in which each parameter enters a single Pauli rotation, the update requires only Pauli measurements on the original and parameter-shifted circuits, without ancillary qubits or controlled operations for overlap estimation.

Three conditions must hold for the approach to succeed. The selected observables must capture the evolution relevant to the target, the circuit must be flexible enough to reproduce their equations of motion, and those equations must be estimated accurately within the measurement budget. These pull against each other: adding observables improves coverage of the target orbit but adds equations for a fixed ansatz to satisfy and quantities to measure.

Our numerics probe the three conditions. Across six six-qubit models with equal shot budgets and ansatz, the observable-targeted update yields longer estimated median times within the target error tolerance than the regularized McLachlan implementation. Among comparisons with observed medians, the ratios range from $1.06$ to $4.22$. Ten-qubit simulations show that this advantage persists for larger systems, even with shallower dictionaries. These results demonstrate the practical potential of observable targeting under finite measurement statistics.

The LiH benchmark exposes the main limitation of the systematic construction of the observable dictionary. With ten qubits, the LiH dictionary at depth $q=2$ already contains $K=2627$ observables and requires about $5\times10^{5}$ measurement settings per velocity evaluation (Table~\ref{tab:numerical-resources}), making each trajectory far more costly than in the lattice models. Because the number of nested commutators grows rapidly with the number of Hamiltonian terms, increasing $q$ quickly becomes impractical for molecular problems. A heuristic construction guided by the structure of the problem offers an alternative. For example, the candidate observables could be restricted using the symmetries and locality of the Hamiltonian, as in symmetry-adapted operator pools \cite{shkolnikov2023avoiding}. Such dictionaries are expected to sustain long reachable times with far fewer observables and measurement settings than a complete commutator expansion.

Several directions remain to be explored. The expansion in Eq.~\eqref{eq:heisenberg-series} suggests weighting the dictionary observables according to their contribution to the target dynamics over the intended time interval. The observable-targeted update is also compatible with improvements developed for standard VQS, such as the adaptive construction of the variational ansatz \cite{yao2021adaptive,linteau2024adaptive,zhang2025adaptive} and measurement reduction, either through more general groupings of compatible observables than the qubit-wise grouping used here \cite{verteletskyi2020measurement, huggins2021efficient, Escudero2023_HEEM, BarberRodrguez2025} or through classical shadows \cite{nakaji2023measurement}. These techniques act on the circuit and on the estimation of its coefficients rather than on the variational principle, and therefore apply equally to both updates. Since the observable-targeted update already yields longer simulation times at equal measurement cost, the savings in circuit depth and measurements it provides are expected to further increase its advantage. Dictionaries could also grow adaptively by testing the equations of motion of additional observables. Further directions include open-system dynamics and implementations in which device noise competes with sampling noise. The broader point is that the observable one intends to measure is information available before the simulation starts, and a variational method need not discard it. 

\emph{Note added.} During the completion of this work, we became aware of related work by A.~A.~Anuar et al.~\cite{anuar2026variational}. Their formulation uses the same least-squares matching of observable equations of motion. Our work focuses on the error in a specified target expectation value and uses an exact error identity to guide the construction of observable dictionaries.

\begin{acknowledgments}
This work was supported by the Government of Spain (Severo Ochoa CEX2019-000910-S, FUNQIP and QEC4QEA-PCI2025-163167), the European Union (PASQuanS2.1, 101113690 and QEC4QEA, 101194322), Fundació Cellex, Fundació Mir-Puig, Generalitat de Catalunya (CERCA program) and the AXA Chair in Quantum Information Science.

The authors used ChatGPT Astra 6 and Claude Opus 5 to assist
with proof development, manuscript drafting
and numerical code implementation. The authors verified
the mathematical arguments, tested the code, and reviewed
the final text, and take full responsibility for the work.
\end{acknowledgments}

\bibliography{bib}

\appendix

\section{Proof of the target-error identity}
\label{app:duhamel}

Let $e(t)=\sigma(t)-\rho(t)$. The definition of the residual and the von Neumann equation imply
\begin{align}
 \dot e(t)&=\mathcal L(e(t))+r(t).
 \label{eq:error-equation-appendix}
\end{align}
With the Schr\"odinger propagator $\Phi_s(X)=\e^{-\ii Hs}X\e^{\ii Hs}$, variation of constants gives
\begin{align}
 e(T)&=\Phi_T(e(0))
       +\int_0^T\Phi_{T-t}(r(t))\,\dd t.
 \label{eq:duhamel-appendix}
\end{align}
Cyclicity of the trace yields
\begin{align}
 \Tr[A\Phi_s(X)]&=\Tr[\Psi_s(A)X].
 \label{eq:duality-appendix}
\end{align}
Taking the trace of Eq.~\eqref{eq:duhamel-appendix} against $A$ therefore gives Eq.~\eqref{eq:target-identity}.

\section{Coefficient estimation and classical updates}
\label{app:implementation}

The coefficients $G$ and $b$ are estimated using Eqs.~\eqref{eq:parameter-shift} and \eqref{eq:force-estimation}, respectively. Pauli strings that commute qubit by qubit can share a measurement setting using single-qubit basis rotations.

The unregularized least-squares solutions satisfy
\begin{align}
 G^\T G u_\star&=G^\T b,
 \label{eq:normal-equations}
\end{align}
where $\T$ denotes transpose. If $G$ has a nontrivial null space, the parameter velocity is not unique. All minimizers give the same rates of change for the dictionary expectations, but may give different rates for observables outside the dictionary.

For $\lambda>0$, the regularized solution is unique and can be written as
\begin{align}
 u_\lambda&=(G^\T G+\lambda I_p)^{-1}G^\T b,
 \label{eq:ridge-solution}
\end{align}
where $I_p$ is the identity on parameter space. In practice, we evaluate this solution using a singular-value decomposition. With finite measurement samples, $G$ and $b$ are replaced by their estimates $\widehat G$ and $\widehat b$.

Both methods use Heun integration. Writing $u(\theta)$ for the velocity computed by the corresponding method, each step takes the form
\begin{align}
 k_1&=u(\theta_n),\nonumber\\
 k_2&=u(\theta_n+\Delta t\,k_1),\nonumber\\
 \theta_{n+1}
 &=\theta_n+\frac{\Delta t}{2}(k_1+k_2).
 \label{eq:heun}
\end{align}
The two velocity evaluations use independent measurement samples, each with a budget of $S$ shots, so a full time step costs $2S$ shots.

\section{Numerical protocol}
\label{app:numerical-protocol}

\subsection{Hamiltonians and initial states}

All models are represented by six and ten qubits. The spin Hamiltonians are
\begin{align}
 H_{\mathrm I}
 &=-J\sum_{j=0}^{4}Z_jZ_{j+1}
   -h_x\sum_{j=0}^{5}X_j
   -h_z\sum_{j=0}^{5}Z_j,\\
 H_{\mathrm{XXZ}}
 &=J_{xy}\sum_{j=0}^{4}
   (X_jX_{j+1}+Y_jY_{j+1})
   \nonumber\\
 &\quad+J_z\sum_{j=0}^{4}Z_jZ_{j+1}
   +h_z\sum_{j=0}^{5}Z_j,\\
 H_{\mathrm D}
 &=J\sum_{j=0}^{4}
   (X_jX_{j+1}+Y_jY_{j+1}+\Delta Z_jZ_{j+1})
   \nonumber\\
 &\quad+\sum_{j=0}^{5}h_jZ_j,\\
 H_{\mathrm{XY}}
 &=\sum_{j=0}^{4}
   (J_xX_jX_{j+1}+J_yY_jY_{j+1})
   -h\sum_{j=0}^{5}Z_j.
\end{align}
For the mixed-field Ising model,
$(J,h_x,h_z)=(1,1.05,0.5)$ and the initial state is
$\lvert000000\rangle$.
For the XXZ model, $(J_{xy},J_z,h_z)=(1,0.5,0.2)$.
The disordered Heisenberg model has $J=\Delta=1$ and
\begin{align}
 (h_0,\ldots,h_5)\simeq
 (&-0.98644,\,0.81105,\,0.96275,\nonumber\\
  &0.55697,\,-0.97929,\,-0.35577),
\end{align}
generated by six uniform draws from $[-1,1]$.
For the XY model, $(J_x,J_y,h)=(1,0.4,0.7)$.
The latter three models start in the N\'eel state
$\lvert101010\rangle$.
The target is $A=Z_2Z_3$ in all four spin models.
The same fields $h_j$ are used for both methods and all measurement budgets. Only the measurement outcomes are resampled between runs.

For ten qubits, the sums extend over ten sites with the same couplings and open boundaries. The Ising chain starts in $\lvert0\rangle^{\otimes10}$, the other three chains start in the N\'eel state $\lvert1010101010\rangle$, and the target is $A=Z_4Z_5$. The disordered fields are
\begin{align}
 (h_0,\ldots,h_9)\simeq
 (&-0.54533,\,-0.36648,\,0.59473,\,0.35251,\nonumber\\
  &-0.21778,\,-0.33437,\,0.19662,\nonumber\\
  &-0.62653,\,0.34551,\,0.88361),
\end{align}
also drawn uniformly from $[-1,1]$.

The three-site Fermi--Hubbard Hamiltonian is
\begin{align}
 H_{\mathrm{Hub}}
 ={}&-t_{\mathrm{hop}}\sum_{j=0}^{1}
       \sum_{\sigma=\uparrow,\downarrow}
       \left(c^\dagger_{j\sigma}c_{j+1,\sigma}
       +c^\dagger_{j+1,\sigma}c_{j\sigma}\right)
       \nonumber\\
 &+U\sum_{j=0}^{2}n_{j\uparrow}n_{j\downarrow},
\end{align}
where $n_{j\sigma}=c^\dagger_{j\sigma}c_{j\sigma}$,
$t_{\mathrm{hop}}=1$, and $U=4$.
Under the Jordan--Wigner transformation, spin-up modes
occupy qubits $0,1,2$ and spin-down modes occupy qubits
$3,4,5$. The initial configuration is
$\lvert\uparrow,\downarrow,\uparrow\rangle$, prepared as
$X_0X_2X_4\lvert0\rangle^{\otimes6}$.
The target is the central double occupancy,
\begin{align}
 A=D_1
 =n_{1\uparrow}n_{1\downarrow}
 =\frac{I-Z_1-Z_4+Z_1Z_4}{4}.
\end{align}
For ten qubits, the chain has five sites, with spin-up modes on qubits $0$--$4$ and spin-down modes on qubits $5$--$9$. The initial configuration is $\lvert\uparrow,\downarrow,\uparrow,\downarrow,\uparrow\rangle$, and the target is the central double occupancy, as for six qubits, 
\begin{align}
    D_2=n_{2\uparrow}n_{2\downarrow}=\frac{I-Z_2-Z_7+Z_2Z_7}{4}.
\end{align}

For LiH, we use the STO-3G integrals at a bond length of $1.45\,\text{\AA}$.
The lowest spatial orbital is frozen as a doubly occupied
core, and spatial orbitals 1--3 form an active space of
two electrons in six spin orbitals.
Relabeling the active spatial orbitals by $p=0,1,2$,
the Jordan--Wigner ordering assigns their spin-up and
spin-down modes to qubits $2p$ and $2p+1$, respectively.
An occupied mode corresponds to $\lvert1\rangle$, with
$n_k=(I-Z_k)/2$.

Both active electrons initially occupy orbital $p=1$,
giving the state $X_2X_3\lvert0\rangle^{\otimes6}$.
The target is its population,
\begin{align}
 A=N_1=n_2+n_3=I-\frac{Z_2+Z_3}{2}.
\end{align}
After omitting the identity contribution, which affects
only the global phase, the molecular Hamiltonian contains
61 Pauli terms.

For ten qubits, spatial orbitals 1--5, all orbitals above the frozen core, form an active space of two electrons in ten spin orbitals. The active orbitals are relabeled as $p=0,\ldots,4$ with the same qubit ordering, the initial state is $X_2X_3\lvert0\rangle^{\otimes10}$, and the target is again $N_1$. The molecular Hamiltonian then contains 275 Pauli terms.

Molecular energies are expressed in Hartree and times in $\hbar/E_{\mathrm h}$. Lattice times are expressed in inverse units of the coupling set to one.

\subsection{Circuits and observable dictionaries}

Each ansatz contains two layers. A layer includes one
rotation $\exp(-\ii\theta_kP_k/2)$ for every distinct
nonidentity Pauli string in the Hamiltonian, with diagonal
strings placed last. Every rotation has an independent
parameter. All parameters are initialized to zero after
preparing the corresponding computational-basis state.

The observable-targeted method uses the commutator construction of
Sec.~\ref{sec:dictionaries} at depth $q=3$ for six qubits and $q=2$ for ten qubits. Each distinct nonidentity Pauli string is included once, with eigenvalues $\pm1$ and equal weight in the objective. The initial sets contain one string for each spin target, three for Hubbard double occupancy, and two for the LiH population. Identity contributions are omitted because
their expectations are constant.

Table~\ref{tab:numerical-resources} gives the numbers of
parameters, dictionary observables, and measurement settings.
A measurement setting is a distinct experimental configuration,
specified by the state-preparation circuit and the measurement
performed on its output. Each setting is repeated for a number
of shots to estimate the required expectation values.
Compatible Pauli observables can share a setting, whereas
different parameter-shifted circuits count as separate settings.

\begin{table}[t]
 \caption{Circuit and measurement resources.
 $p$ is the number of parameters, $K$ is the dictionary
 size at depth $q=3$ for six qubits and $q=2$ for ten qubits, and $N_{\mathrm{tar}}$ and
 $N_{\mathrm M}$ are the numbers of measurement settings
 per velocity evaluation.}
 \label{tab:numerical-resources}
 \centering
 \begin{tabular}{lrrrr}
 \hline\hline
 Model & $p$ & $K$ & $N_{\mathrm{tar}}$ & $N_{\mathrm M}$\\
 \hline
 {\emph{Six qubits}} &&&&\\
 Ising       & 34  & 22  & 697   & 1175\\
 XXZ         & 42  & 115 & 3267  & 1788\\
 Disordered  & 42  & 115 & 3267  & 1788\\
 XY          & 32  & 67  & 2431  & 1043\\
 Hubbard     & 34  & 195 & 3153  & 1176\\
 LiH         & 122 & 414 & 21805 & 14966\\
 \hline
 {\emph{Ten qubits}} &&&&\\
 {Ising}       & {58} & {10}  & {706}  & {1769}\\
 {XXZ}         & {74} & {31}  & {2565} & {2849}\\
 {Disordered}  & {74} & {31}  & {2565} & {2849}\\
 {XY}          & {56} & {23}  & {1945} & {1652}\\
 {Hubbard}     & {62} & {99} & {2306} & {2015}\\
 {LiH}         & {550} & {2627} & {516578} & {152075}\\
 \hline\hline
 \end{tabular}
\end{table}

\subsection{Regularization}
\label{app:calibration}

We regularize both updates to reduce their sensitivity to finite measurement statistics. For a budget of $S$ shots per velocity evaluation, we use the relative strengths
\begin{align}
 \alpha_{\mathrm{tar}}(S)
 &=10^{-3}\sqrt{10^6/S},\nonumber\\
 \alpha_{\mathrm M}(S)
 &=10^{-1}\sqrt{10^6/S}.
 \label{eq:ridge-prescriptions}
\end{align}
These prescriptions are fixed across all models in six qubits. For ten qubits, the observable-targeted prefactor is reduced to $10^{-4}$, so that 
\begin{align}
    \alpha_{\mathrm{tar}}(S)=10^{-4}\sqrt{10^6/S},
\end{align}
while $\alpha_{\mathrm M}$ is unchanged. For the lattice models, the McLachlan prefactor $10^{-1}$ gives the longest medians among $\{10^{-2},10^{-1},1\}$ in every model and budget.

For the observable-targeted method, we use the ridge objective of
Eq.~\eqref{eq:ridge-objective}, with estimated coefficients:
\begin{align}
 u_{\mathrm{tar}}
 &=\underset{u\in\RR^p}{\arg\min}\,
 \left\{\|\widehat G u-\widehat b\|_2^2
 +\lambda_{\mathrm{tar}}\|u\|_2^2\right\},\\
 \lambda_{\mathrm{tar}}
 &=\alpha_{\mathrm{tar}}(S)
   \max\{s_{\max}(\widehat G)^2,10^{-6}\}.
\end{align}
Here, $s_{\max}(\widehat G)$ is the largest singular value
of the estimated derivative matrix. The minimizer is
computed directly using a singular-value decomposition
of $\widehat G$.

For McLachlan evolution, the exact metric is positive
semidefinite, but its estimate can acquire negative
eigenvalues from sampling noise. Writing the symmetric
estimate as $\widehat M=V\operatorname{diag}(\mu_k)V^\T$,
we replace these eigenvalues by zero:
\begin{align}
 \widehat M_+
 &=V\operatorname{diag}(\max\{\mu_k,0\})V^\T.
\end{align}
We then set
\begin{align}
 \lambda_{\mathrm M}
 &=\alpha_{\mathrm M}(S)
   \max\{\lambda_{\max}(\widehat M_+),10^{-6}\}
\end{align}
and solve the regularized system
\begin{align}
 (\widehat M_++\lambda_{\mathrm M}I_p)u_{\mathrm M}
 &=\widehat f.
\end{align}
The floor $10^{-6}$ prevents the regularization from vanishing
when the estimated matrix is zero or very small.

\subsection{Measurement sampling}

We use budgets $S\in\{10^6,10^7,10^8,10^9\}$ per velocity evaluation. Pauli strings are grouped when they commute qubit by qubit. Joint outcomes within each group are sampled from the corresponding multinomial distribution, preserving
correlations between estimators that share measurements. Hadamard-test outcomes in the McLachlan implementation are sampled from binomial distributions.
The simulations include finite measurement statistics but no additional device noise. For $N$ settings, each receives either $\lfloor S/N\rfloor$ or $\lceil S/N\rceil$ shots.

\subsection{Integration and statistical analysis}

Both methods use Eq.~\eqref{eq:heun} with
$\Delta t=5\times10^{-3}$ for six qubits or $\Delta t=2.5\times10^{-3}$ for ten qubits, and $T_{\max}=3$.
Each step uses two independently sampled velocity
evaluations and costs $2S$ shots.
The target expectation is evaluated exactly along
each sampled trajectory and compared with the exact
Hamiltonian evolution. Thus, the diagnostic includes
no additional target-measurement noise.

Target expectations are recorded at intervals of $2.5\times10^{-3}$. The reachable time is the first crossing of the tolerance defined in
Eq.~\eqref{eq:reachable-time}, with $\varepsilon=10^{-3}$ for the lattice models and $\varepsilon=10^{-4}$ for LiH in six qubits, and $\varepsilon=10^{-3}$ for all models with ten qubits.

For each model, method, and budget, we use 100 independent measurement realizations. Trajectories that remain within tolerance through $T_{\max}$ provide only a lower bound on their crossing time. Quantiles that cannot be determined within the simulation window are reported as lower bounds
at $T_{\max}$. The plotted interquartile ranges describe variation among measurement realizations.

\end{document}